\documentstyle[12pt]{article}

\begin{document}

\setcounter{page}{1}

\pagestyle{plain} \vspace{1cm}
\begin{center}
\Large{\bf Effects of the Generalized Uncertainty Principle on the Inflation Parameters}\\
\small \vspace{1cm} {\bf Kourosh Nozari}\quad and\quad {\bf Siamak Akhshabi}   \\
\vspace{0.5cm} {\it Department of Physics,
Faculty of Basic Sciences,\\
University of Mazandaran,\\
P. O. Box 47416-95447, Babolsar, IRAN\\
knozari@umz.ac.ir\\
s.akhshabi@umz.ac.ir}

\end{center}
\vspace{1.5cm}
\begin{abstract}
We investigate the effects of the generalized uncertainty principle
on the inflationary dynamics of the early universe in both standard
and braneworld viewpoint. We choose the Randall-Sundrum II model as
our underlying braneworld scenario. We find that the quantum
gravitational effects lead to a spectral index which is not scale
invariant. Also, the amplitude of density fluctuations is reduced by
increasing the strength of quantum gravitational corrections.
However, the tensor-to-scalar ratio increases by incorporation of
these quantum gravity effects. We outline possible manifestations of
these quantum gravity effects in the recent and future
observations.
\end{abstract}
\vspace{2cm}
\newpage
\section{Introduction}
One of the most important achievements in the spirit of quantum
gravity theories such as string theory, is that the laws of physics
are considerably different at short distances [1-9]. For example,
the standard quantum mechanical commutation relations change to some
modified (generalized) commutators at the length scale of the order
of Planck length [10-11]. These modifications are negligible in low
energy physics but at high energy such as the early universe,
quantum gravitational corrections play a crucial role [12]. As a
result, the standard uncertainty relation of quantum mechanics is
replaced by a gravitational uncertainty relation which contains a
minimal observable length of the order of the Planck length [13].
The very notion of spacetime in this quantum gravity era cannot be
probed more precisely than this minimal observable length [1,2,11].
These fundamental properties of spacetime in quantum gravity era
result in a variety of phenomenologically interesting outcomes for
the rest of physics ( see for instant [14]). In this paper we
investigate the effects of the generalized uncertainty principle on
the inflationary parameters in the standard and braneworld
inflation. The braneworld scenario we analyzed is the
Randall-Sundrum II setup which is one of the most promising
alternatives of extra dimensional theories [15]. One way to discover
the quantum gravitational effects in the inflationary era is to
study the perturbation spectrum generated during inflation. Any
modifications to the scalar and tensorial perturbations spectrum due
to these quantum gravity effects are essentially detectable in the
cosmic microwave background (CMB) data [16-22]. This provides a
reliable approach to test the theories of short distance physics. In
which follows a prime on a quantity denotes differentiation with
respect to its argument.
\section{Inflation and the Generalized uncertainty principle}
In short distances the standard commutation relations will be
changed as [11,13,14]
\begin{equation}
[x_{i},p_{j}]=i\hbar(\delta_{ij}+\beta p^{2}\delta_{ij}+\beta'
p_{i}p_{j}),
\end{equation}
\begin{equation}
[p_{i},p_{j}]=0,
\end{equation}
and
\begin{equation}
[x_{i},x_{j}]=i\hbar\frac{(2\beta-\beta')+(2\beta+\beta')\beta
p^{2}}{(1+\beta p^{2})}(p_{i}x_{j}-p_{j}x_{i}).
\end{equation}
We set $\beta'=0$, so the corresponding Poisson brackets are
\begin{equation}
\{x_{i},p_{j}\}=\delta_{ij}(1+\beta p^{2}),\,\,\,\,\,\,\,
\{p_{i},p_{j}\}=0,\,\,\,\,\,\,\,\{x_{i},x_{j}\}=2\beta
(p_{i}x_{j}-p_{j}x_{i})
\end{equation}
The parameter $\beta$ is related to the minimum length \emph{i.e.}
$x_{min}\sim\sqrt{\beta}$. The calculation of inflationary scalar
density perturbations in the presence of the minimal length are
preformed in Ref. [23]. The scalar field, $\phi$ which drives the
inflation has energy density and pressure
\begin{eqnarray}
\nonumber \rho=\frac{1}{2}\dot{\phi}^{2}+V\\
p=\frac{1}{2}\dot{\phi}^{2}-V
\end{eqnarray}
respectively where $V(\phi)$ is the inflation potential. The
slow-roll parameters are given as usual by [24]
\begin{eqnarray}
\epsilon&=&\frac{M^{2}_{4}}{2}\Big(\frac{V'}{V}\Big)^{2}\\ \nonumber
\eta&=&\frac{M^{2}_{4}}{2} \frac{V''}{V}\,\,.
\end{eqnarray}
In the slow-roll regime we have
\begin{eqnarray}
\nonumber \frac{1}{2}\dot{\phi}^2&\ll& V(\phi)\\ 3H\dot{\phi}
&\simeq& -V'(\phi)
\end{eqnarray}
We now proceed to incorporate the quantum gravitational corrections
as are given by equations (1)-(3) within a typical inflationary
scenario. One could assume that there is a fundamental energy scale
$\Lambda$ (Planck or string scale) that these corrections become
important. Defining the conformal time as [25]
\begin{equation}
\tau=-\frac{1}{aH}\,.
\end{equation}
We will see that physical momentum $p$ and the comoving momentum $k$
are related through
\begin{equation}
k=ap=-\frac{p}{\tau H}\,.
\end{equation}
The conformal time that the new physics rules prior to it is given
by
\begin{equation}
\tau_{0}=-\frac{\Lambda}{Hk}
\end{equation}
where $\Lambda$ is the Planck energy scale. Using equation (1), we
change the comoving momentum $k$ before the time $\tau_{0}$ to
$k(1+\beta k^{2})$. This is a realization of the modified dispersion
relation supported by loop quantum gravity and noncommutative
geometry [26,27]. The equation governing the evolution of
perturbations in the inflation era is [24]
\begin{equation}
\mu''_{k}+\bigg(k^{2}-\frac{a''}{a}\bigg)\mu_{k}=0
\end{equation}
where $\mu$ is a rescaled field $\mu=a\delta\phi$ and a prime
denotes differentiation with respect to $\tau$. The scalar spectral
index in the presence of the minimal length cutoff now is given by
\begin{equation}
n_{s}-1=\frac{d\ln {\cal{P}}_{s}}{d\ln k(1+\beta k^{2})}
\end{equation}
where ${\cal{P}}_{s}$ is the amplitude of the scalar perturbation.
Therefore, we find
\begin{equation}
n_{s}=\frac{1+\beta k^{2}}{1+3\beta k^{2}}\frac{d\ln
{\cal{P}}_{s}}{d\ln k}+1\simeq (1-2\beta k^{2})\frac{d\ln
{\cal{P}}_{s}}{d\ln k}+1\,\,.
\end{equation}
As an important result, the spectral index is not scale invariant in
this case. Any deviation from scale invariance of the spectral index
essentially contains a footprint of these quantum gravity effects.
The change in the Hubble parameter due to the modified commutators
will be realized using slow-roll parameters. At the horizon crossing
epoch we have [24,25]
\begin{equation}
\frac{dH}{dk}=-\frac{\epsilon H}{k}
\end{equation}
changing $k$ to $k(1+\beta k^{2})$ we find
\begin{equation}
H\simeq k^{-\epsilon}e^{-\beta \epsilon k^{2}}\,\,.
\end{equation}
Using equation (11), the tensorial density fluctuation is given by
\begin{equation}
{\cal{P}}_{t}(k)=\frac{1}{a^{2}}<\mid \mu_{k}(\tau)\mid^{2}>
\end{equation}
Following [25], the tensorial density fluctuations in our case is
\begin{equation}
{\cal{P}}_{t}(k)=\bigg(\frac{H}{2\pi}\bigg)^{2}
\bigg(1-\frac{H}{\Lambda}\sin(\frac{2\Lambda}{H})\bigg),
\end{equation}
where the second term on the right hand side is a direct
contribution of quantum gravity effect. For scalar density
fluctuations, one should add an extra $(\frac{H}{\dot{\phi}})^{2}$
term [24,25]
\begin{equation}
{\cal{P}}_{s}(k)=\bigg(\frac{H}{\dot{\phi}}\bigg)^{2}\bigg(\frac{H}{2\pi}\bigg)^{2}
\bigg(1-\frac{H}{\Lambda}\sin(\frac{2\Lambda}{H})\bigg)
\end{equation}
where $H$ is given by equation (15). Figure $1$ shows the
$k$-dependance of tensorial fluctuations for a fixed $\epsilon$ and
$\beta$ while figure $2$ shows the $\beta$-dependance of it for a
fixed $k$. As usual, $\Lambda$ is chosen to be the Planck energy
scale. We note that by variation of $\beta$ ( which is essentially a
fixed quantity related to the minimal observable length) we just
mean a control on the strength of the quantum gravity effect. The
tensor-to-scalar ratio in this setup is given by
\begin{equation}
\frac{{\cal{P}}_{t}}{{\cal{P}}_{s}}=\bigg(\frac{\dot{\phi}}{H}\bigg)^{2}=\Bigg(\frac{16\pi
\sqrt{\epsilon}V}{M_{4}k^{-\epsilon}e^{-\beta \epsilon
k^{2}}}\Bigg)^{2}\,,
\end{equation}
where $M_{4}$ is the $4$-dimensional fundamental scale. The extra
factor $e^{-\beta \epsilon k^{2}}$ in the denominator is the
correction due to the GUP effects. Figure (2) shows the difference
of tensor-to-scalar ratio in standard and modified cases.  As we
see, the tensor-to-scalar ratio increases by incorporation of the
quantum gravity effects.
\begin{figure}[htp]
\begin{center}
\includegraphics{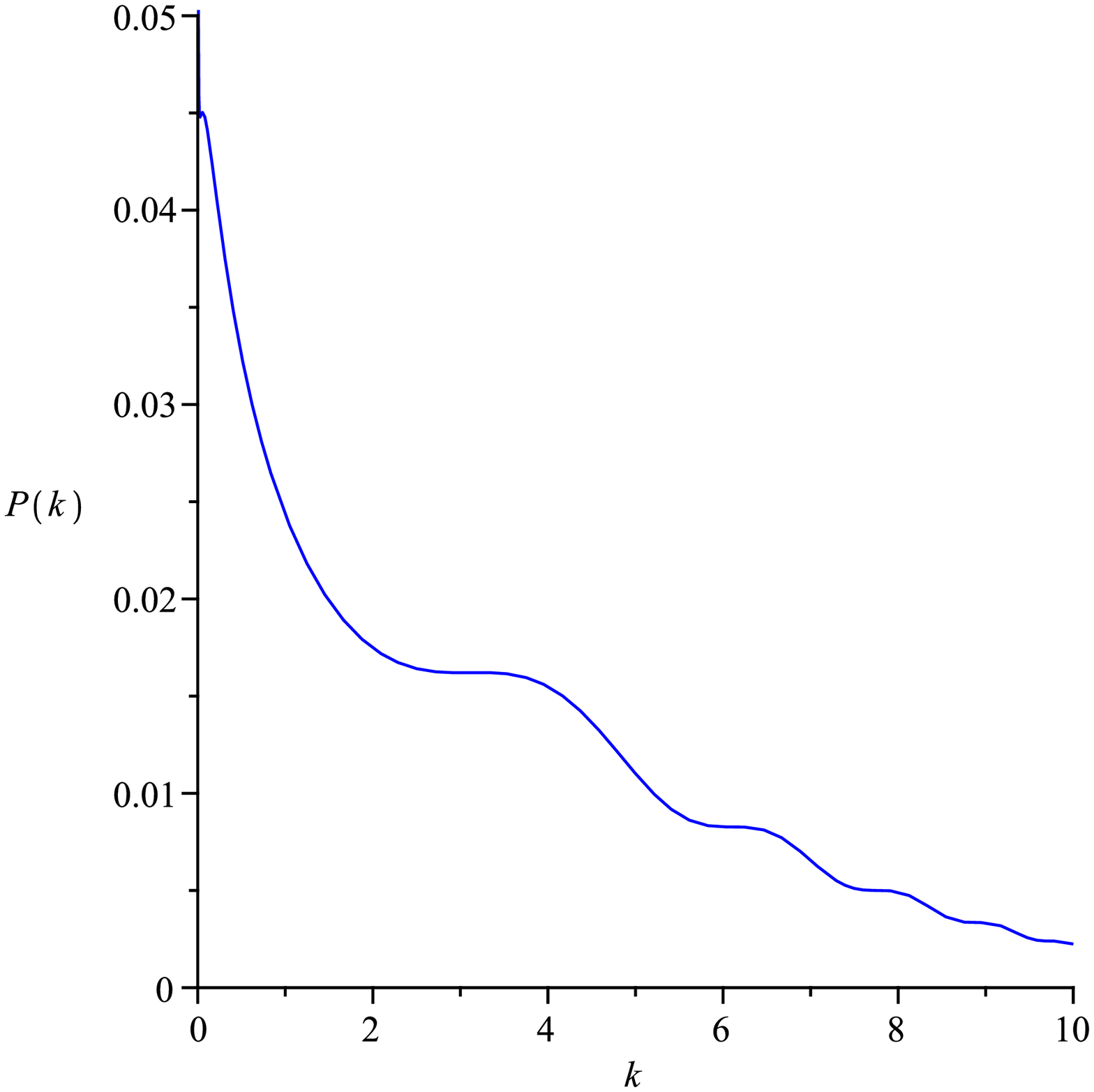} \vspace{5cm}\includegraphics{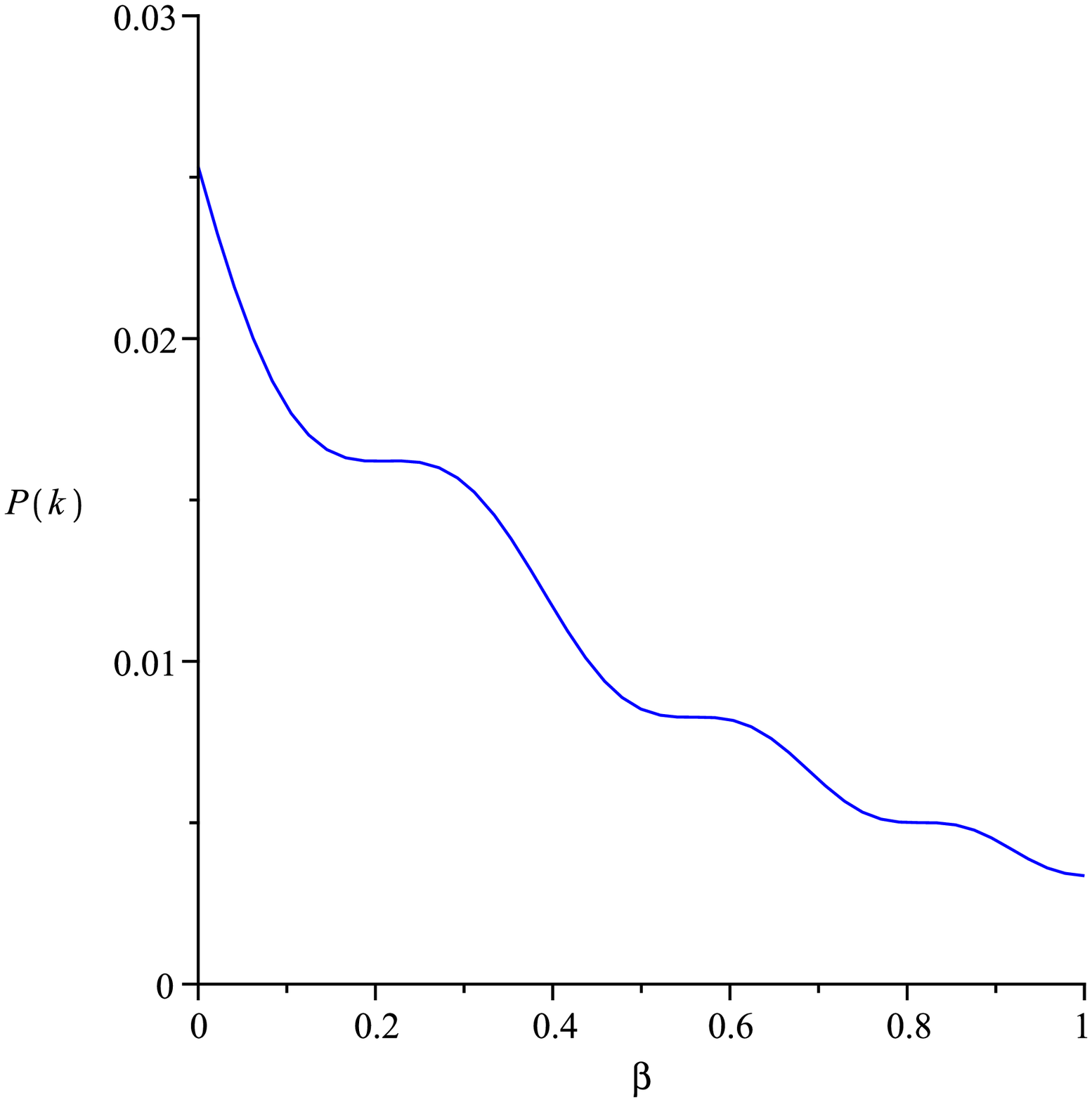}\vspace{2 cm}
\end{center}
\vspace{-0.5cm}
 \caption{\small {a) $k$-dependance of tensorial density
 fluctuations for $\epsilon=\beta=0.1$  b)
 $\beta$-dependance of tensorial density fluctuations
 for  $k=1$. Here there is an oscillatory behavior which can be detected
 essentially in the CMB spectrum as a trace of quantum gravity effects.}}
\end{figure}
\begin{figure}[htp]
\begin{center}\includegraphics{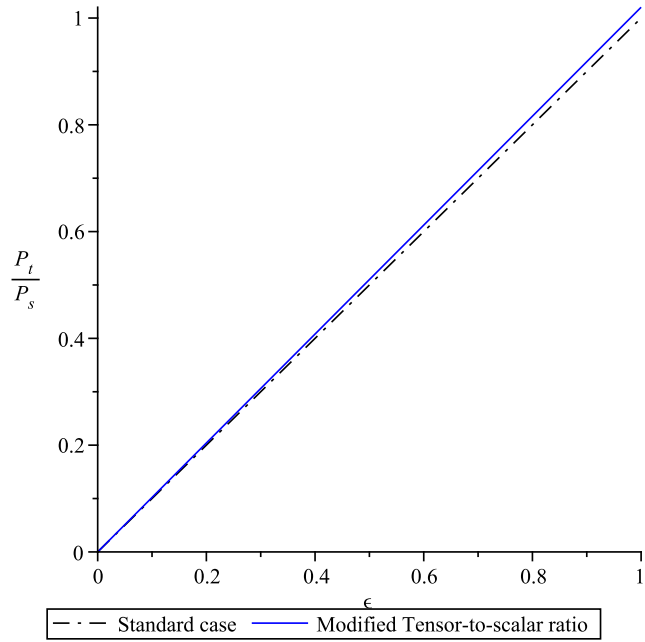} \vspace{1.5cm}
\end{center}
\vspace{6cm}
 \caption{\small {Difference of tensor-to-scalar ratio for standard
 and GUP modified inflation for a fixed $k$ and
 $\beta=0.01$}}
\end{figure}
\newpage
\section{Brane inflation}
Now we turn our attention to a braneworld inflation scenario. We
study the case of Randall-Sundrum II setup [15] in the presence of
the generalized uncertainty principle. The Randall-Sundrum II model
has a single positive tension brane living in an infinite AdS bulk.
Modified Friedmann equation for this setup is given by [28]
\begin{equation}
H^{2}=\frac{\Lambda_{4}}{3}+\bigg(\frac{8\pi}{3M^{2}_{4}}\bigg)\rho+
\bigg(\frac{4\pi}{3M^{3}_{5}}\bigg)^{2}\rho^{2}+\frac{{\cal{E}}}{a^{4}}
\end{equation}
where ${\cal{E}} $ is an integration constant originating from the
projection of the bulk Weyl tensor. The last term is called the dark
radiation and we neglect it in our analysis because during inflation
it will vanish really fast. Also neglecting the cosmological
constant term in the early universe we can rewrite the Friedmann
equation as
\begin{equation}
H^{2}=\frac{8\pi}{3M^{2}_{4}}\rho\bigg[1+\frac{\rho}{2\lambda}\bigg]
\end{equation}
where $\lambda$ is the brane tension. In the inflation era, the only
matter on the brane is a scalar field, $\phi$  with energy density
and pressure defined as (5).

In this braneworld inflation scenario, the slow-roll parameters can
be defined as [29]
\begin{eqnarray}
\nonumber \epsilon&\equiv& \frac{M^{2}_{4}}{16\pi}\bigg(\frac{V'}{V}
\bigg)^{2}\bigg[\frac{2\lambda(2\lambda+2V)}{(2\lambda+V)^{2}}\bigg]\\
\eta&\equiv&
\frac{M^{2}_{4}}{8\pi}\bigg(\frac{V''}{V}\bigg)\bigg[\frac{2\lambda}{2\lambda+V}\bigg]\,.
\end{eqnarray}
The amplitude of scalar perturbations in the slow-roll limit in this
case  is [24,30]
\begin{equation}
{\cal{P}}_{s}= \frac{9}{25}\frac{H^{6}}{{V'}^{2}}.
\end{equation}
 Now we consider the effects of the generalized uncertainty
 principle. Using equations (15) and (23) we find
 \begin{equation}
{\cal{P}}_{s}=\frac{9}{25V'^{2}}\Bigg[k^{-\frac{M^{2}_{4}}{16\pi}\Big(\frac{V'}{V}\Big)^{2}
\Big[\frac{2\lambda(2\lambda+2V)}{(2\lambda+V)^{2}}\Big]}
\exp\bigg\{-\beta\bigg(\frac{M^{2}_{4}}{16\pi}\Big[\frac{V'}{V}\Big]^{2}\Big[\frac{2\lambda(2\lambda+2V)}
{(2\lambda+V)^{2}}\Big]\bigg)k\bigg\}\Bigg]^{6}\,.
 \end{equation}
We choose the inflaton field potential to be chaotic type
\emph{i.e.} $V(\phi)=\frac{1}{2}m^{2}\phi^{2}$ so equation (24)
becomes
\begin{equation}
{\cal{P}}_{s}=\frac{9}{25m^{4}\phi^{2}}\Bigg[k^{-\frac{M^{2}_{4}}{4\pi\phi^{2}}
\bigg[\frac{2\lambda(2\lambda+m^{2}\phi^{2})}{(2\lambda+
\frac{1}{2}m^{2}\phi^{2})^{2}}\bigg]}\times
e^{-\beta\bigg({\frac{M^{2}_{4}}{4\pi\phi^{2}}
\bigg[\frac{2\lambda(2\lambda+m^{2}\phi^{2})}{(2\lambda+
\frac{1}{2}m^{2}\phi^{2})^{2}}\bigg]}\bigg)k}\Bigg]^6
\end{equation}
Figure (2a) shows the scalar density perturbations against $k$ for
given values of other parameters while figure (2b) shows its
behavior in terms of $\beta$. We see that relative to standard case,
${\cal{P}}_{s}$ reduces by incorporation of quantum gravity effects.
Also, ${\cal{P}}_{s}$ reduces by enhancement of the role played by
quantum gravity effects.
\begin{figure}[htp]
\begin{center}
\includegraphics{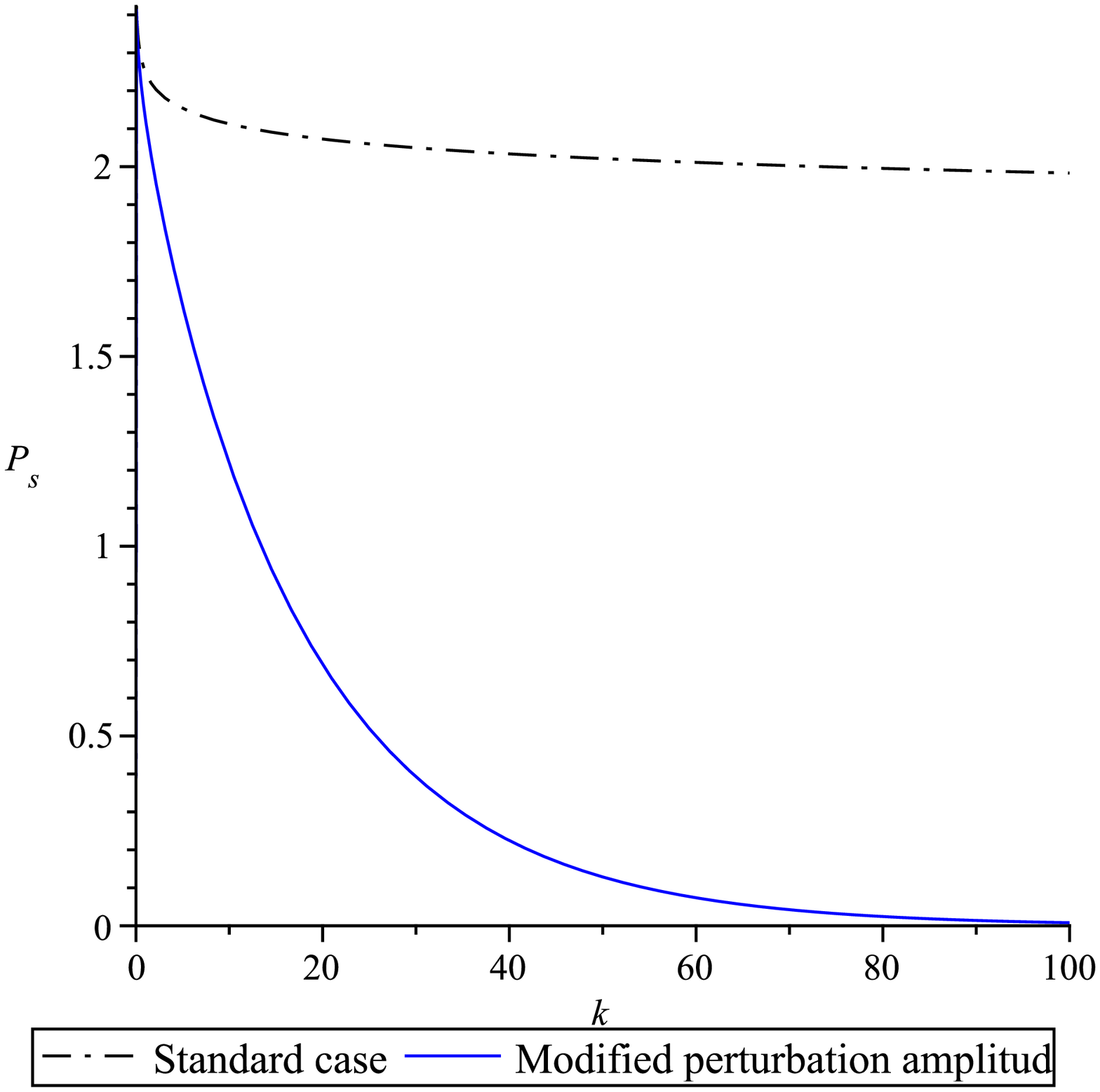} \vspace{7cm}\includegraphics{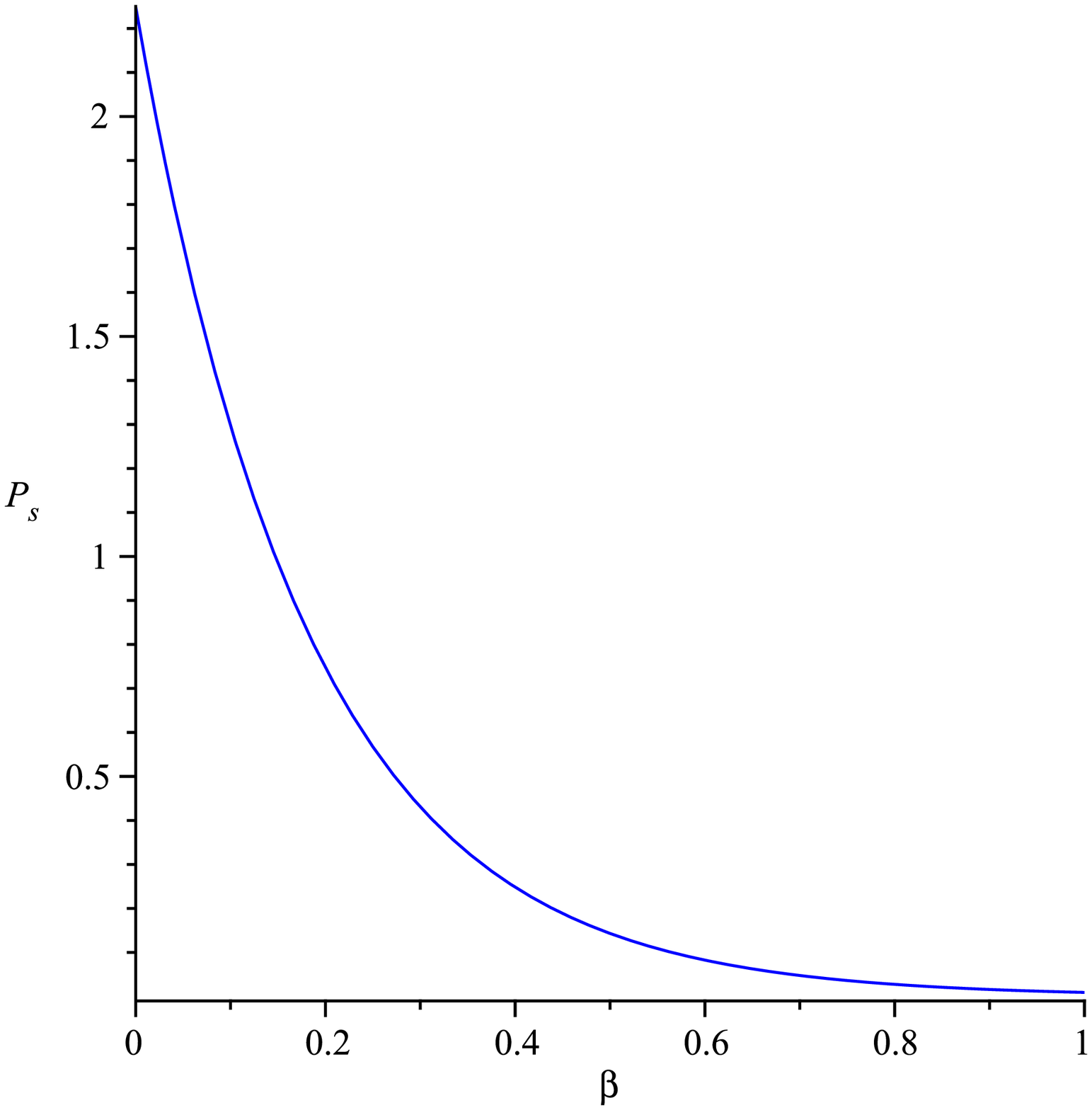}\vspace{4 cm}
\end{center}
\vspace{-4cm} \caption{\small { a) $k$-dependance of scalar density
fluctuations for $\beta=0.01$, $M_{4}=1$, $m=0.1M_{4}$,
$\lambda=0.1M_{4}$\, and\, $\phi=4M_{4}$.  b) $\beta$-dependance of
scalar density fluctuations for  $k=1$, $M_{4}=1$, $m=0.1M_{4}$,
$\lambda=0.1M_{4}$\,and\, $\phi=4M_{4}$.  }}
\end{figure}

Now the scalar spectral index is
\begin{equation}
n_{s}-1\equiv\frac{d\ln {\cal{P}}_{s}}{dk}
\end{equation}
Using equation (25) we plot the spectral index against $k$ and
$\beta$. Figure $3$ shows the result of this calculation. If the
quantum gravity effects are relatively small, the spectral index
shows a scale invariance behavior. However, in the limit of strong
quantum gravity effects, the spectral index has no scale invariance
characteristics.
\begin{figure}[htp]
\begin{center}
\includegraphics{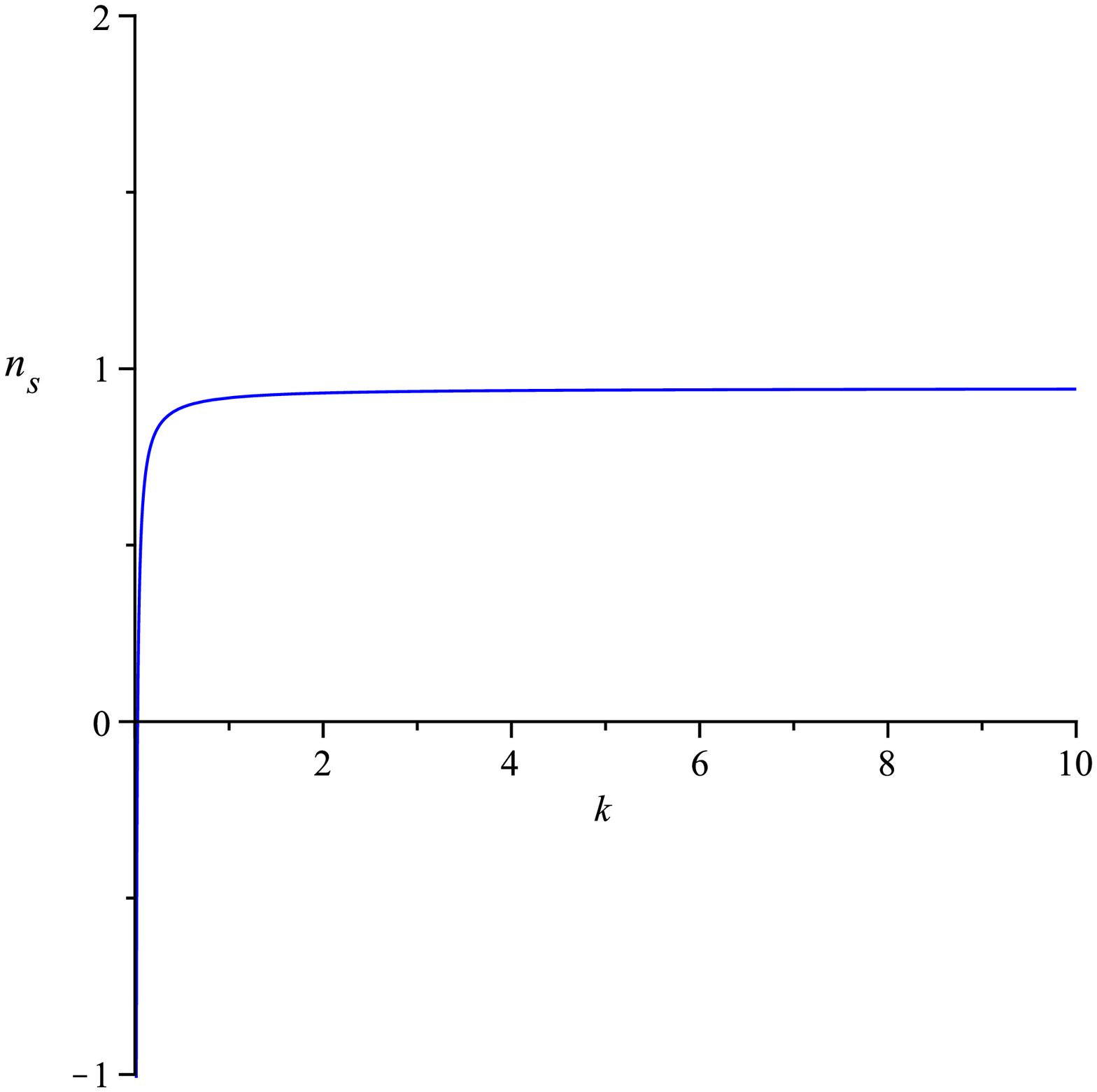} \vspace{5cm}\includegraphics{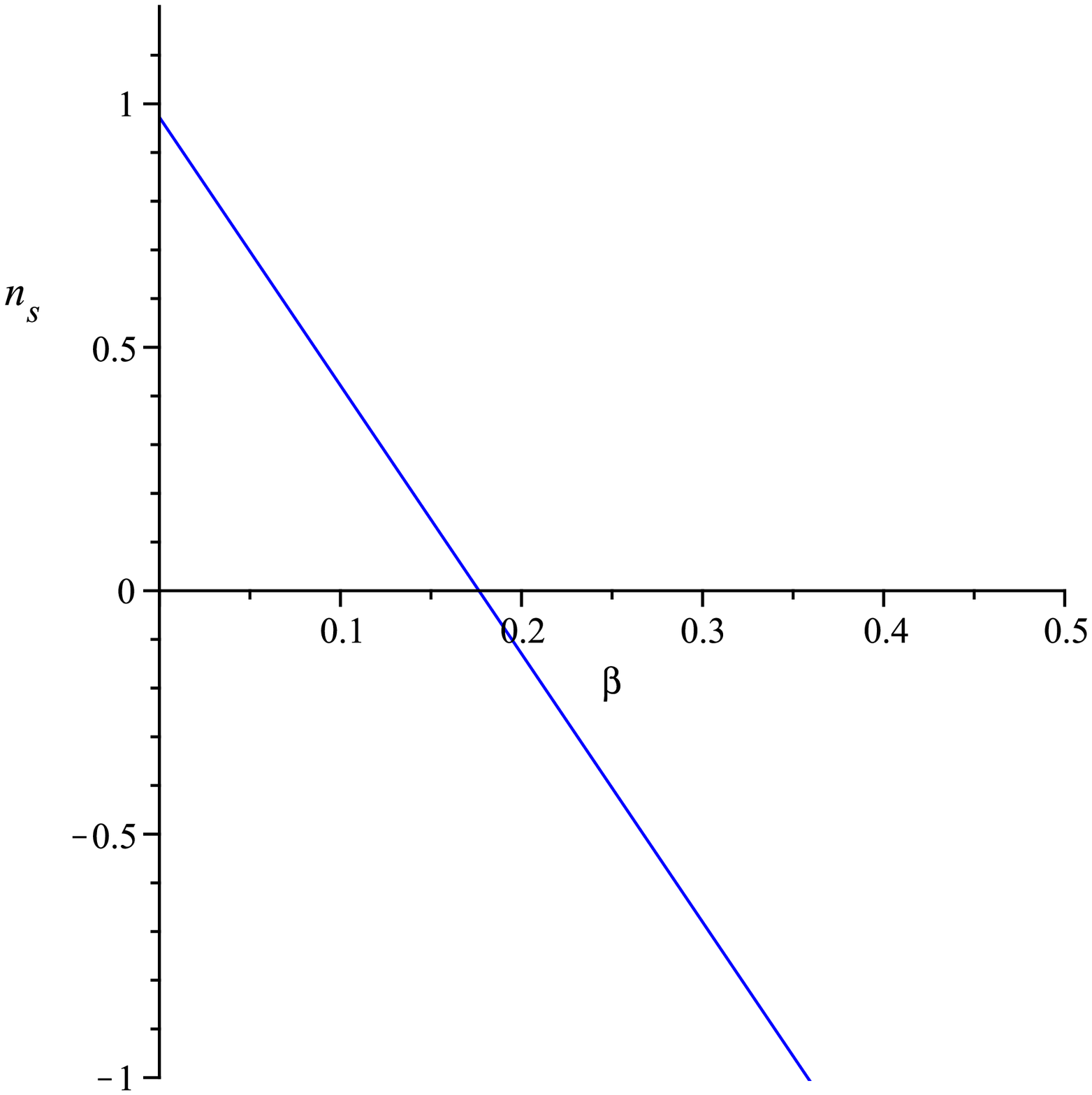}\vspace{2 cm}
\end{center}
\vspace{0cm} \caption{\small {a)  $k$-dependance of the scalar
spectral index for $\beta=0.01$, $M_{4}=1$, $m=0.1M_{4}$,
$\lambda=0.1M_{4}$\, and \,  $\phi=4M_{4}$.   b)  $\beta$-dependance
of the scalar spectral index for  $k=1$, $M_{4}=1$, $m=0.1M_{4}$,
$\lambda=0.1M_{4}$\, and \, $\phi=4M_{4}$.  }}
\end{figure}

\newpage
\section{Conclusion}
In this letter we have considered the effects of the generalized
uncertainty principle ( as a common feature of all promising quantum
gravity candidates ) on the inflationary dynamics of both the
standard $4D$ theory and the Randall-Sundrum II braneworld setup. As
an important result, we have shown that in the presence of the
strong quantum gravity effects, the spectral index is not scale
invariant. In this sense, any deviation from scale invariance of the
spectral index essentially contains a footprint of these high energy
effects. There is an oscillatory behavior in the $k$-dependance of
density fluctuations which can be detected essentially in the CMB
spectrum as a trace of these effects. The tensor-to-scalar ratio
increases by incorporation of the quantum gravity effects. In a
braneworld viewpoint of inflation on the Randall-Sundrum II brane,
we have shown that relative to the standard case, ${\cal{P}}_{s}$
reduces by incorporation of the generalized uncertainty principle.
Also, ${\cal{P}}_{s}$ reduces by enhancement of the role played by
quantum gravity effects. In this case, similar to $4D$ case, only in
the limit of week quantum gravity effects the spectral index shows a
scale invariant behavior. However, in the limit of strong quantum
gravity effects, the spectral index has no scale invariance
properties. We note that any modifications to the scalar and
tensorial perturbations spectrum (such as oscillatory behavior in
the amplitude of fluctuations ) due to these quantum gravity effects
are essentially detectable in the spectrum of the cosmic microwave
background radiation and this may provide an indirect test of
quantum gravity proposals.

\end{document}